\documentclass[apj]{emulateapj}
\usepackage{hyperref}
\newcommand{\msun}{{\rm\ M}_\odot}

\usepackage[T1]{fontenc}

\shorttitle{Morpho-kinematics of Nova Mon 2012}
\shortauthors{Ribeiro, Munari \& Valisa}

\begin{document}


\title{Optical Morphology, Inclination and Expansion Velocity of the \\ Ejected Shell of Nova Monocerotis 2012}

\author{V. A. R. M. Ribeiro\altaffilmark{1}}
\affil{Astrophysics, Cosmology and Gravity Centre, Department of Astronomy, University of\\ Cape Town, Private Bag X3, Rondebosch 7701, South Africa}
\email{vribeiro@ast.uct.ac.za}

\author{U. Munari}
\affil{INAF Astronomical Observatory of Padova, 36012 Asiago (VI), Italy}
\email{ulisse.munari@oapd.inaf.it}

\and

\author{P. Valisa}
\affil{ANS Collaboration, c/o Osservatorio Astronomico, via dell'Osservatorio 8, 36012 Asiago (VI), Italy}
\email{paolo.valisa@gmail.com}

\altaffiltext{1}{South African Square Kilometer Array Fellow.}

\begin{abstract}
The morphology of the ejected shell of the He/N Nova Monocerotis 2012 outburst was studied in detail. Synthetic line profile spectra were compared to the [O~{\sc iii}] 4959,5007~\AA\ emission line profiles in order to find the best fit morphology, inclination angle and maximum expansion velocity of the ejected shell. The simplest morphology was found to be that of a bipolar structure with an inclination angle of 82$\pm6$~degrees and a maximum expansion velocity of 2400$^{+300}_{-200}$~km/s (at day 130 after outburst). Such a high degree of shaping is un-expected for a system with a main sequence star (as suspected from the systems colors). The degree of shaping may be disentangled with resolved optical imaging. Furthermore, these results may be confirmed with radio imaging which is expected to follow the same gross features of the outburst as the optical band and the high inclination implied here can be corroborated with a 7.1 hour period which has been suggested to arise from partial eclipses of extended emission by an accretion disk rim.
\end{abstract}


\section{Introduction}
A thermonuclear classical nova outburst occurs on the surface of a White Dwarf (WD, the primary) following extensive accretion from a less evolved companion star (the secondary, in most cases a main sequence star). Another class of thermonuclear runaway are those which show multiple outbursts in their recorded history, named recurrent novae \citep[here the secondary is, mostly, an evolved star,][]{DRB12}. The outburst ejects 10$^{-5}$ -- 10$^{-4}$ $\msun$ of matter at velocities of order hundreds to thousands of kilometers per second \citep[e.g.][]{BE08,B10}. The short recurrence time in recurrent novae has been attributed to a high mass WD, probably close to the Chandrasekhar limit, together with a high accretion rate \citep{SST85,YPS05}.

Nova Monocerotis 2012 was first detected by the {\it Fermi} Gamma-ray Large Area Telescope as a $\gamma$-ray source on 2012 June 22 \citep[Fermi J0639+0548,][]{CHV12}, taken as $t_0$. At the time of the {\it Fermi} observations, Nova Mon 2012 was not known to be a bona fide nova in outburst, due to its proximity to the Sun, until independent discovery by Shigehisa Fujikawa on 2012 August 9.8 UT of a possible nova at a magnitude of 9.4 \citep{FYN12}. It was subsequently confirmed as a nova spectroscopically and associated with the $\gamma$-ray source \citep{CSD12}. The photometric evolution of Nova Mon 2012 was discussed by Munari et al. (2013). High resolution spectroscopy obtained on August 20 by \citet{MDV12} showed unblended interstellar lines corresponding to an E$(B-V)$ = 0.30 and a distance $>$ 1~kpc from the Sun. Two distance estimates for this object were provided in the literature, ranging from around 1.4~kpc \citep{CCN12} to 3 -- 4~kpc \citep{CSD12}, all consistent with an intervening Interstellar Medium located 1~kpc \citep{MDV12}, implying a distance for the nova greater than 1~kpc. On 2012 August 19, {\it Swift}/XRT observations showed a hard and absorbed X-ray spectrum, with the majority of the counts at energies above 2~keV \citep{NMC12}. Nova Mon 2012 was also observed in the radio with the VLA \citep{CCN12}, resolved as a double radio source with the e-VLBI on 2012 September 18 \citep{OYP12} and with the OVRO 40m, Effelsberg 100m and IRAM 30m telescopes \citep{FRB12}. The e-VLBI observations showed two compact components aligned North West -- South East and separated by about 35~mas, implying an expansion velocity of 0.4 mas/day \citep{OYP12}.
\begin{table*}
\caption{Log of spectroscopic observations. The third column is days after outburst (from 2012 June 22).}\label{tab1}
\centering
\begin{tabular}{crrcccc}
\hline
\multicolumn{2}{c}{Date}&
%\multicolumn{1}{c}{}&
\multicolumn{1}{c}{$\Delta t$}&
\multicolumn{1}{c}{Exp. Time}&
\multicolumn{1}{c}{Resolution}&
\multicolumn{1}{c}{$\lambda$ range}&
\multicolumn{1}{c}{Telescope}\\
\multicolumn{1}{c}{2012}&
\multicolumn{1}{c}{UT}&
\multicolumn{1}{c}{(days)}&
\multicolumn{1}{c}{(sec)}&
\multicolumn{1}{c}{}&
\multicolumn{1}{c}{(\AA)}&
\multicolumn{1}{c}{}\\
\hline\hline
Aug 20 &  3.250  &  60 & 1800   & 11,000  & 4200$-$8650 & 0.6m+ECH \\
Aug 28 &  3.283  &  68 & 4200   & 17,000  & 4200$-$8650 & 0.6m+ECH \\
Sep 05 &  3.418  &  76 & 2400   & 23,000  & 3670$-$7335 & 1.8m+ECH \\
Sep 08 &  3.417  &  79 & 4500   & 17,000  & 4100$-$8550 & 0.6m+ECH \\
Oct 24 &  0.667  & 125 & 3600   & 17,000  & 4100$-$8650 & 0.6m+ECH \\
Oct 29 & 23.783  & 130 & 1200   & 16,000  & 3670$-$7335 & 1.8m+ECH \\
Nov 20 & 22.500  & 152 & 4500   & 11,000  & 4200$-$8650 & 0.6m+ECH \\
Dec 28 & 22.283  & 190 & 1900   & 16,000  & 3670$-$7335 & 1.8m+ECH \\
\hline
\end{tabular}\end{table*}
\begin{figure*}
\plotone{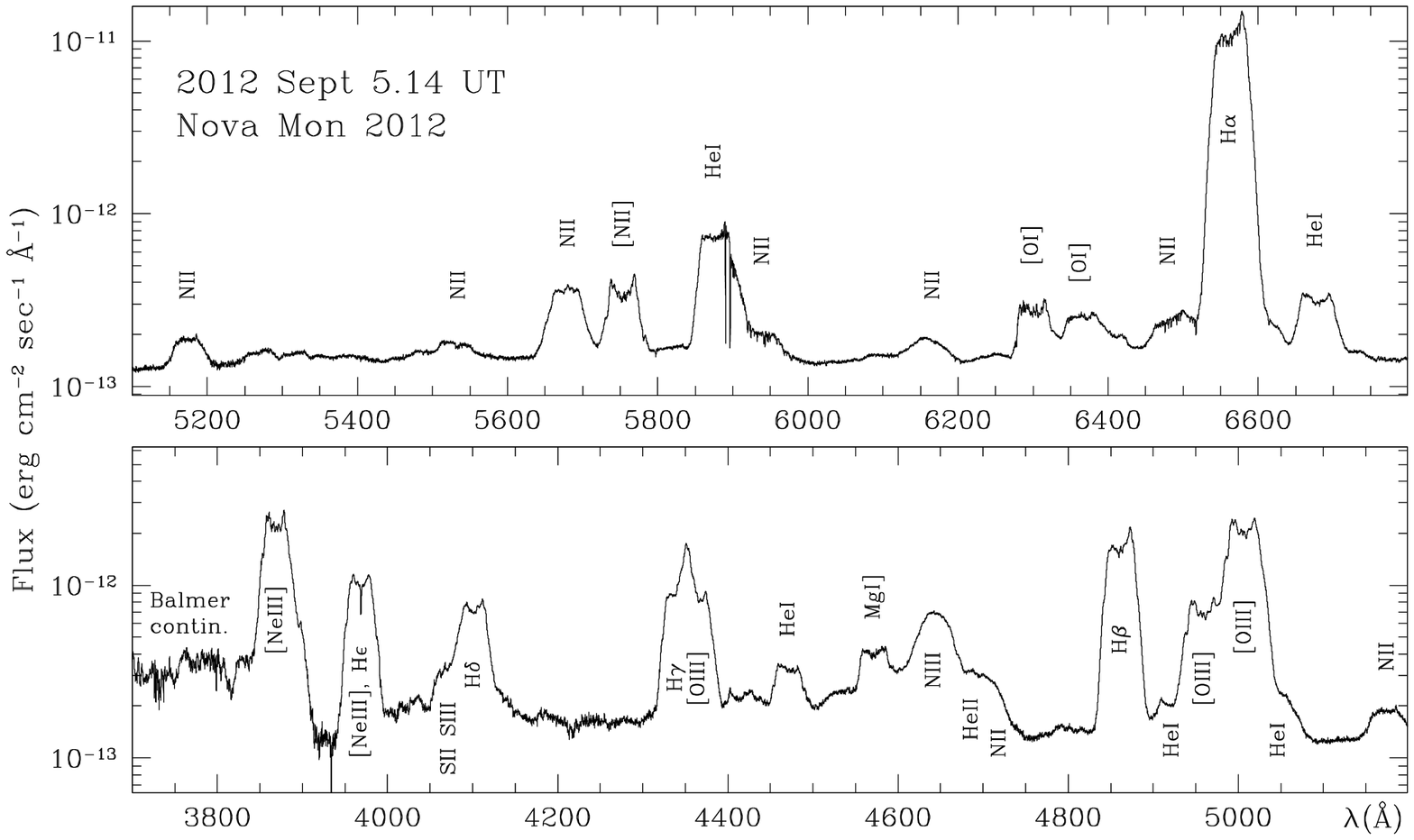}
\caption{Identification of emission lines in the spectrum of Nova Mon 2012 for 2012 Sept 5.14 UT. The ordinates are logarithm of the absolute flux to emphasize visibility of weak features. Note the sharp interstellar atomic absorption lines by Ca~{\sc ii} 3933,3967~\AA\ and Na~{\sc i} 5890,5896~\AA. \label{fig1}}
\end{figure*}

\citet{GDS12} reported on a likely progenitor from IPHAS and a coincident source from the UKIDSS \citep{LHL08}, UGPS J069978.60+055332.9 at $J$~=~16.26, $H$~=~15.71 and $K$~=~15.42. For E$(B-V)$~=~0.3 and $d$~=~1~kpc, the corresponding absolute magnitude is $M_K$~=~4.9, as expected for a K5~V star \citep[mass = 0.67~$\msun$,][]{DL00}.  If an accretion disk was contributing significantly to the infrared (IR) flux, the donor star would be even cooler and less massive. No significant contribution from an accretion disk is confirmed by the \citet{GDS12} IPHAS colors, ($r^{\prime} - i^{\prime}$)~=~0.69 or ($r^{\prime} - i^{\prime}$)~=~0.44 after correction for reddening, which places the object to the red side of the K5~V color, ($r^{\prime} - i^{\prime}$)~=~0.36. Adopting intrinsic IR colors from \citet{SL07} and reddening relation from \citet{FM03}, a K5~V star would have a $(J-K)$~=~0.88, close to the observed $(J-K)$~=~0.84.

This object will provide great details for the study of nova remnant shaping and, due to its suggested small distance, has the potential to become one of the best studied novae. In this paper, we aim to disentangle the morphology of the remnant by means of morpho-kinematical studies of the [O~{\sc iii}] 4959,5007~\AA\ emission line profiles. With this information we may determine the inclination angle and maximum expansion velocity of the system.

The use of emission line profiles to disentangle such information has been extensively applied to other novae \citep[e.g.][]{S83, GO99,GO00,RBD09,RDB11,MRB11,R11}. Although morpho-kinematical studies do not directly provide the formation mechanism we can infer such mechanisms by analyzing the retrieved morphology. In section~\ref{obs}, our spectroscopic observations are presented while in section~\ref{model} we discuss our modeling techniques and associated assumptions. Section~\ref{results} shows our model fits to the observed line profiles. In section~\ref{discussion} we discuss our results and finally present the conclusions in section~\ref{conclusions}.

\section{Observations}\label{obs}
High resolution (ranging from 11,000 to 23,000), high S/N Echelle spectra of Nova Mon 2012 were obtained with the REOSC Echelle spectrograph attached to the 1.82m telescope operated in Asiago by INAF Astronomical Observatory of Padova, and the Multi-Mode Spectrograph mounted on the 0.6m telescope at Campo dei Fiori (Varese).  The multi-order Echelle spectra were absolutely flux calibrated against observations of the standard star HR~1578, located nearby on the sky, observed immediately before or after the nova, and then merged into a 1D continuous spectra. The log of the observations is shown in Table~\ref{tab1} and in Figure~\ref{fig1} the low resolution spectrum for 2012 September 5.14 UT is shown to demonstrate the general spectral appearance of the wider wavelength coverage. The low resolution spectrum was obtained with the Asiago 1.22m B\&C telescope of the University of Padova, at a dispersion of 2.31 \AA/pix. This spectrum demonstrates the object to be a He/N nova, as defined by \citet{W92} and recent spectral developments indicate Nova Mon 2012 to be a Neon nova implying a massive WD \citep{M13}.

\section{Modeling}\label{model}
Optical imaging of resolved nova shells has shown a myriad of structures \citep[e.g.][]{H72,S83,SOD95,GO00,HO03}. Several mechanisms for the formation of these structures were put forward. For example, a common envelope phase during outburst, the presence of a magnetized WD and an asymmetric thermonuclear runaway \citep[see e.g.][]{OB08}. The common envelope phase is the most widely accepted formation mechanism, where the ejecta engulfs the secondary star within a matter of minutes following the outburst. The secondary then transfers energy and angular momentum to the ejecta \citep{LSB90,LOB97,POB98}. When the WD rotation is incorporated into the calculations of the common envelope phase it produces the observed prolate remnants \citep{POB98}. Recent smoothed particle hydrodynamic calculations show that the mass loss from the secondary, during quiescence, is highly concentrated in the orbital plane and that this produces naturally the bipolar structures of the ejecta, with possibly an equatorial waist \citep{MP12,MBP13}.
\begin{figure}
\plotone{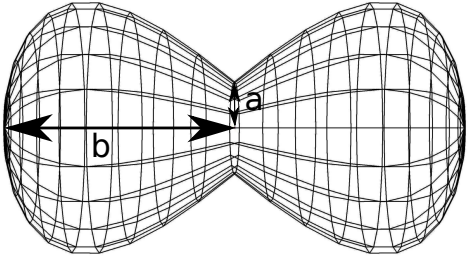}
\caption{Nova Mon 2012 model structure as visualized in {\sc shape}. The inclination of the system is defined as the angle between the plane of the sky and the central binary system's orbital plane, as observed if the page were the plane of the sky. The letters a and b represent the semi-minor and -major axis, respectively, of the ejected material. The ratio of the semi-minor to semi-major axes defines the squeeze, see Eq.~\ref{eq1}. \label{fig2}}
\end{figure}

Morpho-kinematical modeling involves the disentanglement of the morphology and the kinematics of an object. Several studies, mentioned above, have in one way or another used this technique, although they were also aided with resolved imaging, to constrain the models further. Work by \citet{SOD95}, later updated by \citet{B02}, showed what appears to be a relationship between the speed class of the nova and the major to minor axis ratio of the expanding nova shell. Here, the faster the nova speed class the less the degree of shaping. However, this relationship appears not to apply for systems such as RS Ophiuchi which showed a deprojected major to minor axial ratio of 3.85 \citep{RBD09}. This asymmetry may be due to the fact that in RS Oph there is a pre-existing red-giant wind and therefore a different formation mechanism is at play \citep[e.g.][]{BHO07,SRM08,RBD09}.
\begin{figure*}
\resizebox{\hsize}{!}{\includegraphics{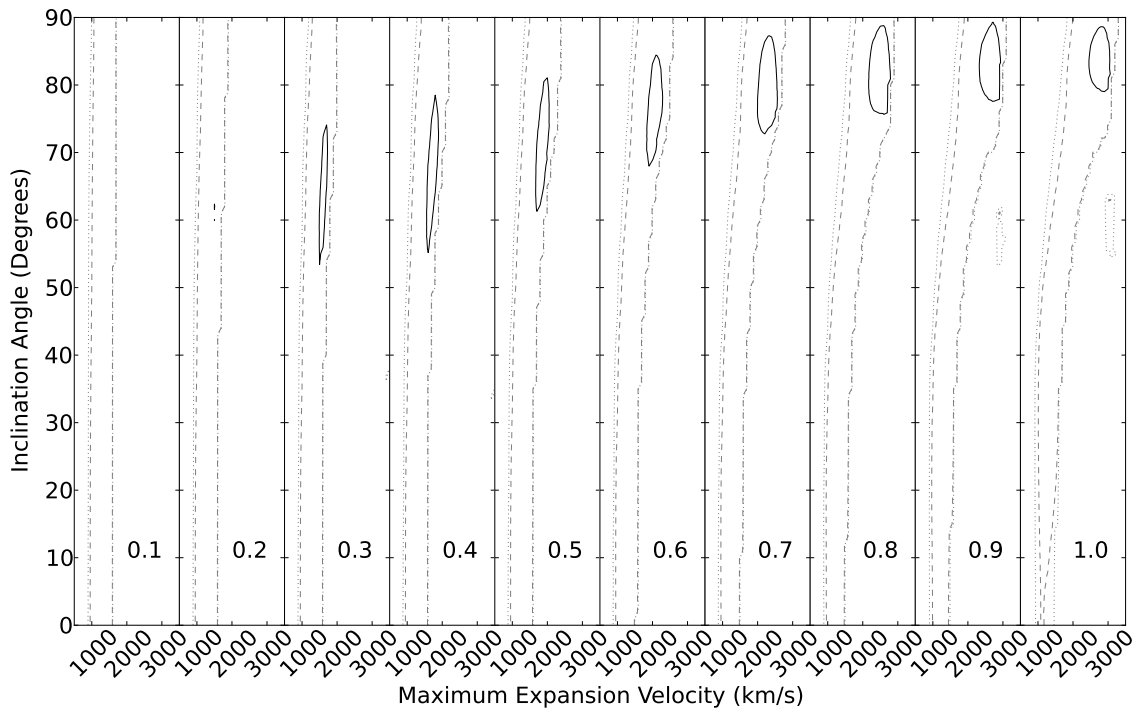}}
\caption{P--value distribution of the synthetic spectra fit to the day 130 after outburst spectrum. The contour lines represent different values of the PDF, 1$\sigma$ (solid black), 0.05 (dashed gray) and 0.01 (dotted gray) for each squeeze (inset numbers on lower right of each sub-plot. The best fit result was that of a squeeze of 0.8, maximum expansion velocity of 2400$^{+300}_{-200}$~km/s and inclination angle of 82$\pm6$~degrees (see also Table~\ref{tab2}). The best fit results for each squeeze are also shown in Figure~\ref{fig4}.\label{fig3}}
\end{figure*}

Several assumptions have gone into building the model morphologies described above, first and foremost they are based on observed morphologies \citep[e.g.][]{H72,S83,SOD95,GO00,HO03,RBD09}. In \citet{MRB11} and \citet{RDB11}, several models were constructed, using {\sc shape}\footnote{Available from \url{http://bufadora.astrosen.unam.mx/shape/index.html}} \citep{SKW11}, for detailed fits of the early outburst spectra of V2672 Oph and V2491 Cygni, respectively. This amounted to 43680 individual synthetic spectra for different morphologies, inclination angles from 0 to 90 degrees, and maximum expansion velocities from 100 to 8000~km/s (any higher and we would be reaching the supernova regime). In the extremely fast V2672 Oph, \citet{MRB11} observed the first 8 days following outburst while for V2491 Cyg, \citet{RDB11} investigated a larger time-span. To fit the spectra for V2491 Cyg, \citet{RDB11} assumed that the change in the H$\alpha$ line profile arose due to the termination of the post-outburst wind phase and complete ejection of the envelope, an interpretation put forward for V382 Velorum \citep{VPD02}. This interpretation aided greatly in finding the best fit structure at later times, replicating the observed line profiles with a combination of H$\alpha$ and [N~{\sc ii}], both arising from the same morphology.
\begin{table}
\caption{Results of the $\chi^2$ / degree of freedom (dof) fits to day 130 after outburst spectrum. Each synthetic spectrum $\chi^2$ / dof is provided with its respective P--value. The errors were determined for P--values greater than 1$\sigma$, with the exception of the Squeeze = 0.1 where the P--value fell below 1$\sigma$ (in this case a P--value of 0.05 was taken). \label{tab2}}
\centering
\begin{tabular}{lccccc}
\hline
Squeeze & V$_{\textrm{exp}}$ & Inclination & $\chi^2$ / & P-value \\
 & (km/s) & (Degrees) & dof & \\
\hline\hline
0.1 & 1500$^{+100}_{-500}$ & 58$^{+32}_{-58}$ & 1.799620 & 0.615018 \\
0.2 & 1500$^{+0}_{-0}$ & 62$^{+0}_{-2}$ & 1.483159 & 0.686163 \\
0.3 & 1600$^{+100}_{-100}$ & 62$^{+12}_{-8}$ & 1.130754 & 0.769656 \\
0.4 & 1700$^{+100}_{-100}$ & 68$^{+10}_{-12}$ & 1.005915 & 0.799821 \\
0.5 & 1800$^{+200}_{-100}$ & 70$^{+11}_{-8}$ & 0.973879 & 0.807572 \\
0.6 & 2100$^{+100}_{-200}$ & 80$^{+4}_{-12}$ & 0.893796 & 0.826925 \\
0.7 & 2300$^{+200}_{-300}$ & 80$^{+7}_{-7}$ & 0.601627 & 0.896060 \\
0.8 & 2400$^{+300}_{-200}$ & 82$^{+6}_{-6}$ & 0.574938 & 0.902145 \\
0.9 & 2600$^{+300}_{-200}$ & 83$^{+6}_{-5}$ & 0.739620 & 0.863848 \\
1.0 & 2700$^{+300}_{-200}$ & 83$^{+5}_{-3}$ & 0.842072 & 0.839380 \\
\hline
\end{tabular}\end{table}

\citet{R11} and \citet{RBD13} investigated a new technique which automatically improves the fits of a set of parameters using a least squares minimization technique. For a qualitative study two tests were performed: (i) varying the inclination angle (in steps of 1$^{\circ}$ from 0 -- 90 degrees, where an inclination $i$ = 90$^{\circ}$ corresponds to the orbital plane being edge-on, and $i$ = 0$^{\circ}$ being face-on) and maximum expansion velocity (in steps of 100~km/s from 100 -- 8000 km/s) to retrieve synthetic emission line profiles and then a $\chi^2$ test performed on the results to find the best fit parameters. Note that the maximum expansion velocity is at the apex of the shell. Then (ii) using the inbuilt optimization technique, the inclination angle and maximum expansion velocity were allowed to vary. The outcome was that the optimization module produced results within the one sigma uncertainty when compared to the results retrieved with the former method \citep{R11,RBD13}. The latter test allows for the quick exploration of different geometries and then applying the former technique a model can be better constrained and uncertainties derived (which at the moment of writing this paper, the latter technique does not allow for).

The modeling performed here is a departure from previous work where the H$\alpha$ emission line profile was modeled, either due to the lack of unblended forbidden lines or lack of forbidden lines altogether since these are optically thin. Although, due to the fact that structure is observed in the line profile of H$\alpha$, which suggests that part of the line comes from optically thin regions, H$\alpha$ suffers some self absorption which may explain some of the observed asymmetry. Furthermore, from photoionization modeling H$\alpha$ comes from the whole of the ionized ejecta, whereas [O~{\sc iii}] comes from the outer regions, which are the most important and sensitive to the expansion velocity and geometry of the ejecta.

We first performed a $\chi^2$ test on the model grids applied to day 8.33 and day 25 after outburst for V2672 Oph and V2491 Cyg, respectively. The grid consisted of the following morphologies; i) polar blobs with an equatorial ring, ii) a dumbbell structure with an hour-glass over-density, iii) a prolate spheroid with an equatorial ring, iv) a prolate spheroid with tropical rings and v) a prolate spheroid with polar blobs and an equatorial ring \citep[from][]{MRB11,RDB11}. These were applied to the observed spectrum on 2012 October 29, primarily as a consistency test, as none of these models should fit the late time spectra because they are all based on the early time assumption before the termination of the post-outburst wind phase and complete ejection of the envelope \citep{VPD02}. For example, this was an important factor when considering the spectra of V2491 Cyg, where the same geometry was applied to the different evolutionary stages of the spectra \citep{RDB11}. Visual inspection of the synthetic spectrum with the lowest $\chi^2$ showed that these did not replicate the observed spectrum.
\begin{figure}
%\epsscale{0.5}
\plotone{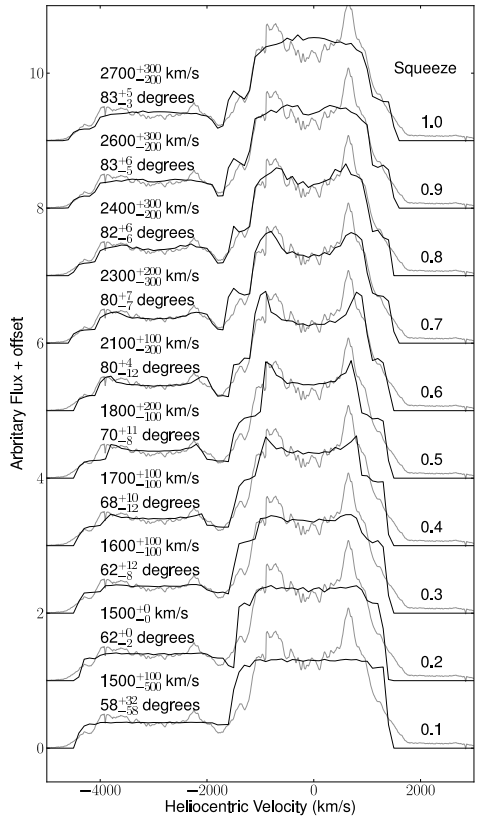}
\caption{Best fit synthetic spectra (black) to Nova Mon 2012 spectrum on day 130 after outburst (gray, centered on the [O~{\sc iii}] 5007~\AA\ emission line). On the right-hand side is the squeeze, one minus the ratio between the semi-minor to the semi-major axes of the ejected shell to the center of the binary system, and on the left the best fit maximum expansion velocity and inclination angle (with their respective errors). The ratio of the semi-minor to semi-major axes defines the squeeze, see Eq.~\ref{eq1}.\label{fig4}}
\end{figure}
\begin{figure}
%\epsscale{0.5}
\plotone{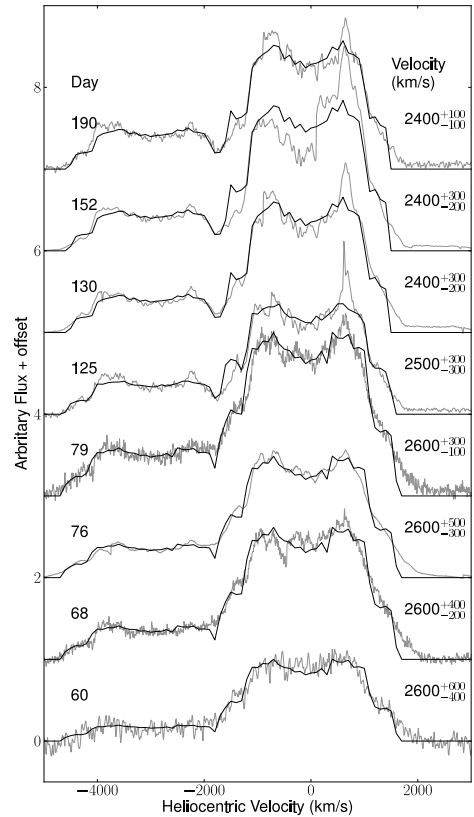}
\caption{Best fit synthetic spectrum (black) to Nova Mon 2012 spectrum (gray, centered on the [O~{\sc iii}] 5007~\AA\ emission line) spectra on the respective days after outburst (inset left) for a bipolar morphology with a squeeze of 0.8 and inclination angle of 82 degrees as fixed parameters and only varying the maximum expansion velocity (inset right-hand side are the best fit values). The ratio of the semi-minor to semi-major axes defines the squeeze, see Eq.~\ref{eq1}. \label{fig5}}
\end{figure}

Without prior knowledge of the system morphology, different structures were built. These included, polar blobs, a prolate structure with an equatorial ring, a prolate structure with tropical rings and bipolar structures. To replicate complete ejection of the envelope our morphologies were surface distributed (equivalent to a thin shell). We first appied the optimizer technique in order to constrain a geometry. We found this to be a bipolar structure. The full parameter space for inclination, maximum expansion velocity and degree of shaping (squeeze, defined below) were then explored to retrieve the synthetic emission line spectrum to compare with the observed spectra to determine $\chi^2$. The squeeze (a modifier within {\sc shape}) compresses or expands a structure as a function of position along the symmetry axis (in this case the major axis). We use the fractional amount by which the object is compressed to the squeeze axis (in this case the minor axis). In its simplest form we define the squeeze as\footnote{for fuller discussion and examples on the squeeze modifier use, see \url{http://bufadora.astrosen.unam.mx/shape/v4/manual/v4.0/Tutorials\_Written/Module\_3D/Modifiers/3D\_modifiers\_squeeze.html}}:
\begin{equation}
\textrm{squeeze} = 1 - \frac{\textrm{a}}{\textrm{b}}
\label{eq1}
\end{equation}
where a and b are the semi-minor and semi-major axes of the ejected shell distance to the center of the binary system (Figure~\ref{fig2}). In other words, the degree by how much the waist of the ejecta is pinched. Hence a squeeze of 0.0 would represent a sphere.

\section{Results}\label{results}
Synthetic spectra were fitted to the observed spectrum on day 130 after outburst (Table~\ref{tab2}, Figures~\ref{fig3} and \ref{fig4}). In Table~\ref{tab2}, the best fit results for the maximum expansion velocity and inclination angle along with their respective $\chi^2$ / degree of freedom (dof) values and the P--value distribution are shown. The errors were determined assuming a P--value greater than 1$\sigma$. The best fit parameter (assuming the highest P--value) is found for a system with a squeeze of 0.8 (Figure~\ref{fig2}), maximum expansion velocity of 2400$^{+300}_{-200}$~km/s and inclination angle of 82$\pm6$~degrees. Furthermore, shown in Figure~\ref{fig3} are the results for the P-values as a contour plot, for three P-values, 1$\sigma$, 0.05 and 0.01.

For a visual comparison of the goodness of fit from Table~\ref{tab2}, the best fit results for individual synthetic spectra are shown in Figure~\ref{fig4} comparing to the day 130 after outburst spectrum. What is evident is that the best fit synthetic spectra are those with a high value of squeeze, replicating well the general features of the observed spectrum. It is worth to note that we cannot replicate all the high velocity material due to the fact that the models have a well defined ``edge''. Although, as shown in Figure~\ref{fig1}, there is also some contribution from He~{\sc i} in the blue and red wings of [O~{\sc iii}] 4959,5007~\AA, respectively. Furthermore, in performing these fits we found that the spectrum was displaced by +80 km/s from a null baricentric velocity, which we associate with the systemic recessional velocity of the system.

The best fit parameters were applied to the remainder of the spectra in order to retrieve their expansion velocities (Figure~\ref{fig5}). Keeping the squeeze and inclination angle the same as the result on day 130 after outburst. It is sufficient to say that most of the spectra are well reproduced by a squeeze of 0.8 and inclination angle of 82 degrees. As mentioned before, we are unable to replicate much of the wings of the observed spectra due to our well defined ``edge'' in the models. The actual edge of the visible ejecta is a dynamical concept during the outburst evolution. During the initial, optically thick evolution, as the ionized fraction of the ejecta grows with time, so does the range of observed velocities widen. Once the ejecta turns optically thin and completely ionized, the width of the emission lines is seen to shrink because the rate of recombination in the outer, faster moving ejecta declines as $r^{-3}$.

\section{Discussion}\label{discussion}
Living with just the data at our disposal, the bipolar morphology is the simplest fit we were able to converge on. Refined model assumptions and fitting constraints will grow in pace with the availability, in time, of more data, and wider wavelength coverage. A bipolar origin of the morphology is also consistent with the e-VLBI observations of two components which may be associated with the ejecta \citep{OYP12}. We should expect the radio and optical to follow the same gross features of the outburst. For example, \citet{BHO07} suggested the size of the structure in the RS Oph remnant to be consistent with the expected expansion of the emitting regions imaged earlier at radio wavelengths \citep{OBP06}.

Interpreting the origin of just the bipolarity, without prior knowledge of the progenitor system, leaves some open questions as to its origin. The infrared colors would suggest that the system has a main sequence star, according to the prescription of \citet{DRB12}. If this assumption is correct the degree of shaping is much larger than that expected from \citet{SOD95} and \citet{B02}. Furthermore, understanding the origin of the bipolar morphology may become clearer with an interpretation of the origin of the $\gamma$-ray emission. At least in one case of $\gamma$-ray emission in novae the system is known to be a symbiotic \citep[V407 Cyg, e.g.][]{MJA11}.

\begin{figure}
\plotone{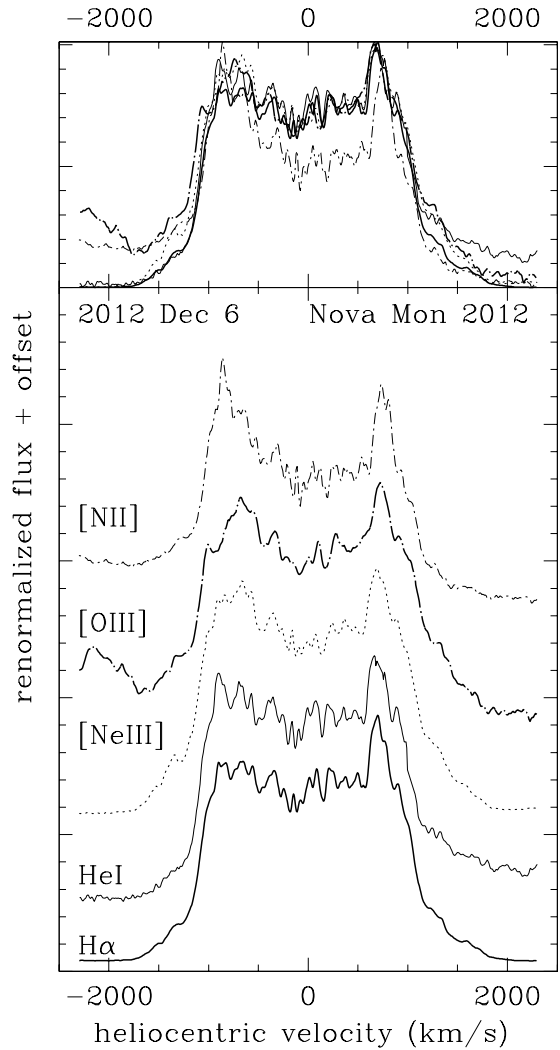}
\caption{Comparison of forbidden lines in Nova Mon 2012 with the permitted lines of H$\alpha$ and He~{\sc i} on day 130 after outburst, showing very similar line profiles implying the geometry is the same, even though arising from different parts of the gross ejecta. \label{fig6}}
\end{figure}

It is noteworthy that most lines in Figure~\ref{fig1} have the same general profile except for N~{\sc iii} 4640, which has a Gaussian shape (resembling a filled sphere), the permitted N~{\sc ii} with a flat top (not unlike a thin shell sphere) and the double peaked forbidden [N~{\sc ii}] (as a bipolar structure). Even if some degree of blending exists here with the underlying lines (for example [Ne~{\sc iv}] 4721), we interpret this as arising from different regions, for example the forbidden line must form in the outer ejected material while the N~{\sc iii} probably originates from a denser interior. In Figure~\ref{fig6} the line profiles for some of the forbidden lines and He~{\sc i} and H$\alpha$ are shown, displaying very similar appearances which trace the same overall geometry; however, the saddle-like profiles of nebular lines suggests they form mainly in the outer ``skin'' of the the expanding bipolar shells, while the less pronounced saddle-like profile of permitted lines suggest part of their emission is contributed by the ``body'' of the bipolar shells. The residual asymmetry in the permitted lines is most likely due to a combination of self-absorption in the ejecta and/or clumpy material.

\section{Conclusions}\label{conclusions}
We argue that the optical line profiles are well replicated with a bipolar morphology (squeeze = 0.8), a high inclination angle ($\sim$ 82$\pm6$ degrees, i.e. the major axis laying almost on the plane of the sky) and maximum expansion velocity of $\sim$ 2400$^{+300}_{-200}$ km/s on day 130 after outburst. When the model is applied to spectra on other dates, the implied velocities are within the errors derived for the day 130; however, the poor fit to some of the earlier spectra may suggest high velocity material that quickly fades as the overall ejecta expands. Furthermore, such a high inclination suggest that we should observe eclipses from this system. In fact, these results are corroborated by the observations of a persistent 7.1 hour orbital period, supposedly orbital in origin \citep{OBP13,WWS13,MCD13}, suggested to be due to partial eclipses of extended emission by an accretion disk rim \citep{OBP13}. If the progenitor system contains a main sequence star we need to revise our understanding of the shaping mechanism in these systems, as a common envelope phase followed by transfer of energy and angular momentum from the main sequence star would not replicate this degree of shaping of the ejecta.

Clues to the origin of the geometry may be ingrained in the origin of the $\gamma$--ray emission. For example, in V407 Cyg the $\gamma$--ray emission is thought to arise from interaction of the ejecta with the heavy mass losing Mira variable companion to the outbursting WD \citep{AAA10,MJA11}, affecting the shaping as in RS Oph \citep[e.g.,][]{OBP06,BHO07}. Nova Mon 2012 was detected prior to outburst with the Infrared Survey Explorer (WISE) in all four bands \citep{BJA13}. The color comparison suggested a large mid-IR excess most likely associated with the dusty extended envelope of a companion star. \citet{BJA13}  suggested that the $\gamma$-ray emission may have a similar emission mechanism as in V407 Cyg. Further work is required in this respect though, and in particular understanding the origin of this IR excess.

\acknowledgments
The authors would like to thank W. Steffen and N. Koning for valuable discussions on the use of {\sc shape} and A. Milani, J. Kos and M. Zerjal for some assistance with the observations. The South African SKA Project is acknowledged for funding the postdoctoral fellowship position at the University of Cape Town. We thank an anonymous referee for valuable and insightful comments on the original manuscript.

{\it Facilities:} \facility{Asiago 1.82m}, \facility{Varese 0.6m}.

%\bibliographystyle{apj} % style aa.bst
%\bibliography{../../reference} % your references Yourfile.bib

\end{document}